\documentclass[aip,amsmath,amssymb,reprint]{revtex4-1}
\pdfoutput=1 
\usepackage[utf8]{inputenc}
\usepackage[T1]{fontenc}
\usepackage{mathptmx}
\usepackage{amsmath}
\usepackage{amsfonts}
\usepackage{graphicx}
\usepackage{dcolumn}
\usepackage{bm}
\usepackage{float}
\usepackage{threeparttable}
\usepackage{upquote}
\usepackage{multirow}
\usepackage{hyperref}
\usepackage{xcolor}
\usepackage{units}

\def\beq{\begin{equation}}
\def\eeq{\end{equation}}
\def\bea{\begin{eqnarray}}
\def\eea{\end{eqnarray}}
\def\brcl{\begin{array}{rcl}}
\def\bccl{\begin{array}{ccl}}
\def\blcl{\begin{array}{lcl}}
\def\err{\end{array}}

\begin{document}

\title{Wasserstein metric for improved QML with adjacency matrix representations}

\author{Onur \c{C}aylak}
\affiliation{Department of Mathematics and Computer Science, Eindhoven University of Technology, P.O. Box 513, 5600 MB Eindhoven, The Netherlands}
\affiliation{Institute for Complex Molecular Systems, Eindhoven University of Technology, P.O. Box 513, 5600 MB Eindhoven, The Netherlands}
\affiliation{Institute for Pure and Applied Mathematics, University of California Los Angeles, 460 Portola Plaza, Los Angeles, California 90095, United States}

\author{O. Anatole von Lilienfeld}
\email{anatole.vonlilienfeld@unibas.ch}
\affiliation{Institute of Physical Chemistry and National Center for Computational Design and Discovery of Novel Materials (MARVEL), Department of Chemistry, University of Basel, 4056 Basel, Switzerland}
\affiliation{Institute for Pure and Applied Mathematics, University of California Los Angeles, 460 Portola Plaza, Los Angeles, California 90095, United States}

\author{Bj\"orn Baumeier}
\email{b.baumeier@tue.nl}
\affiliation{Department of Mathematics and Computer Science, Eindhoven University of Technology, P.O. Box 513, 5600 MB Eindhoven, The Netherlands}
\affiliation{Institute for Complex Molecular Systems, Eindhoven University of Technology, P.O. Box 513, 5600 MB Eindhoven, The Netherlands}
\date{\today}

\begin{abstract}
We study the Wasserstein metric to measure distances between molecules represented by the atom index dependent adjacency ``Coulomb'' matrix, used in kernel ridge regression based supervised learning. Resulting quantum machine learning models exhibit improved training efficiency and result in smoother predictions of molecular distortions. We first demonstrate smoothness for the continuous extraction of an atom from some organic molecule. Learning curves, quantifying the decay of the atomization energy's prediction error as a function of training set size, have been obtained for tens of thousands of organic molecules drawn from the QM9 data set. In comparison to conventionally used metrics ($L_1$ and $L_2$ norm), our numerical results indicate systematic improvement in terms of learning curve off-set for random as well as sorted (by norms of row) atom indexing in Coulomb matrices. Our findings suggest that this metric corresponds to a favorable similarity measure which introduces index-invariance in any kernel based model relying on adjacency matrix representations.
\end{abstract}

\maketitle

\section{Introduction}
The application of machine learning (ML) to atomistic simulation has  been gaining traction over recent years~\cite{Behler_NNreview2015,RuppForeword2015, BehlerPerspective2016, rupp2018guest,noe2019machine,musil2019machine,ceriotti2019unsupervised,ceriotti2019machine,Anatole2019exploring}. Kernel ridge regression (KRR) models of quantum properties (Q) applicable throughout chemical compound space (CCS) was established in 2012~\cite{CM}, and has been growing ever since~\cite{Montavon_NIPS2012,Montavon2013MachineSpace,Hansen2013AssessmentEnergies,ML4Polymers_Rampi2013,Sandip2016}. See Refs.~\cite{CeriottiScienceUnified2017,QMLessayAnatole} for more details and references. By now, QML has become a viable and popular tool for generating surrogate property models enabling rapid estimates of relevant molecular and materials properties, holding great promise for computational materials and molecular design~\cite{fpdesign2014anatole,marzari2016materials}, as recently exemplified for the discovery of nearly ninety stable crystal candidates in the Elpasolite structure~\cite{Elpasolite_2016}. 

When setting up standard KRR based QML models of some quantum property $P$ (aka ``label'')~\cite{Hansen2013AssessmentEnergies}, three fundamental choices must be made, (i) the representation $\mathbf{x}$ (aka ``feature''), (ii) the kernel function ($k$), and (iii) the metric ($\text{dist}(\cdot,\cdot)$), such that 
\bea
P(\mathbf{x}) & = & \sum_i^{N} \beta_i k(\text{dist}(\mathbf{x,x}_i)),
\eea
where $N$ and $\{\beta_i\}$ correspond to number and regression coefficients of training instances, respectively.
The representation of a chemical system is known to play an important role. For example, when using incomplete representations (or non-unique), proof was given that QML models can produce absurd results~\cite{FourierDescriptor}. While the details of the representation (other than uniqueness) are less crucial for artificial neural networks, the specific definition of how a chemical system is being specified is known to dramatically affect the learning efficiency of KRR based QML models. Namely, encoding the right physics, such as translational or atom-index invariance, in the representation results in systematic reduction of quantum data needs for achieving the same pre-defined predictive accuracy~\cite{BAML}. This is of particular interest since QML models are typically trained within scarce data regimes due to (a) the immense computational (or experimental) cost for generating labels and (b) the tremendous scale of CCS~\cite{ChemicalSpace, anatole-ijqc2013}. Due to its obvious impact on QML model performance, it is not surprising that substantial efforts have been made to improve upon the representations. For example, using atomization energies of organic molecules stored in the QM9 dataset~\cite{QM9}, various benchmark results have been obtained including as representations the Coulomb matrix and BOB (2015)~\cite{BOB}, BAML (2016)~\cite{BAML}, HDAD (2017)~\cite{HDAD}, constant-size-descriptors (2018)~\cite{collins2018constant}, SLATM~\cite{Amons} (2017), SOAP (2017)~\cite{CeriottiScienceUnified2017}, FCHL (2018)~\cite{FCHL}, MBD (2018)~\cite{pronobis2018many}, and wavelets (2018)~\cite{eickenberg2018solid}). See Ref.~\cite{Anatole2019exploring} for a joint graphical illustration of learning curves coming from these, as well as artificial neural network based, models. Apart from the representation, the choice of kernel function is also known to affect the performance of the QML model, as shown in Refs.~\cite{Rabitz1996,Hansen2013AssessmentEnergies, soloviov2015reproducing,unke2017toolkit,CeriottiScienceUnified2017}. Within this paper, we focus on the aforementioned third choice: The metric. More specifically, previous studies have predominantly relied on  Euclidean or Manhattan norms as a metric. This choice can be questioned when it comes to atom index dependence of adjacency matrices, such as the Coulomb matrix (CM)~\cite{CM}, possibly resulting in discontinuities in the surrogate model due to displacement (or alchemical change~\cite{anatole-ijqc2013}) of the nucleus. In this paper, we discuss how such spurious artefacts are resolved by using a more sophisticated, distribution based measure: The Wasserstein metric~\cite{Wasserstein1969}.

\section{Method}
The CM is an adjacency matrix with diagonal and off-diagonal terms corresponding to approximate free atom and nuclear repulsion contributions to the total potential energy of a molecule, respectively~\cite{CM}. Its adaptation to crystal representations was published subsequently~\cite{MLcrystals_Felix2015}. Its creation was motivated by the fact that it is unique for fixed molecular charges up to permutation of atoms, and that first-principles calculations also require only nuclear coordinates $\mathbf{R}_I$ and nuclear charges $Z_I$ as input. 
\begin{equation}
    C_{IJ} = \begin{cases}
    0.5 Z_I^{2.4} & I=J\\
    \frac{Z_IZ_J}{\vert \mathbf{R}_I - \mathbf{R}_J\vert} & I\ne J\\
    \end{cases}.
\end{equation}
As the CM is invariant to 3D translations and rotations of a molecule, it intrinsically ensures that the molecule's potential energy is constant under those transformations. Among the early problems identified for the general application of the Coulomb matrix is the index dependence. Sorting the Coulomb matrix such that $\vert\vert C_m \vert\vert \geq \vert\vert C_{m+1} \vert\vert \,$ for all $m$, where $C_m$ is the $m$-th row, renders the CM bijective up to rotation and translation. While appealing for its simplicity and still in use for various applications~\cite{Caylak2019EvolutionaryMaterials,ghosh2019deep}, the use of the sorted CM must be cautioned when applied to situations in which smooth geometrical (or alchemical) changes are under consideration. Such applications include training throughout CCS and subsequent  prediction of energies in molecular dynamics trajectories, or geometry relaxations energies, when sorting can lead to a swapping of indices in the vectorized forms of the CM, and sudden atomic index reassignments between test and query system. This {\em indexing problem} manifests itself in discontinuities in predictions and/or the need for a large number of data points for training the respective models. 

Here, rather than attempting to resolve the indexing problem through ever more sophisticated representations, for example using atom centered symmetry functions~\cite{Neuralnetworks_Behler2011}, SOAP~\cite{BartokGabor_Descriptors2013}, HDAD~\cite{HDAD}, SLATM~\cite{Amons}, MBD~\cite{pronobis2018many}, or FCHL~\cite{FCHL}, we investigate if this issue can also be resolved by using a different metric, capable of alleviating the sudden re-assignment occurring within $L_1$ or $L_2$ norms. In particular, the Wasserstein norm of order 1, which is denoted by $W_1$, is a natural way to compare two probability distributions $p$ and $q$~\cite{Wasserstein1969}. This norm is widely used in various fields, like machine learning, image processing, and signal processing~\cite{Wasserstein15, montavon2015wasserstein, rabin2014adaptive}. 

\begin{figure}
\centering
\includegraphics[width=\linewidth]{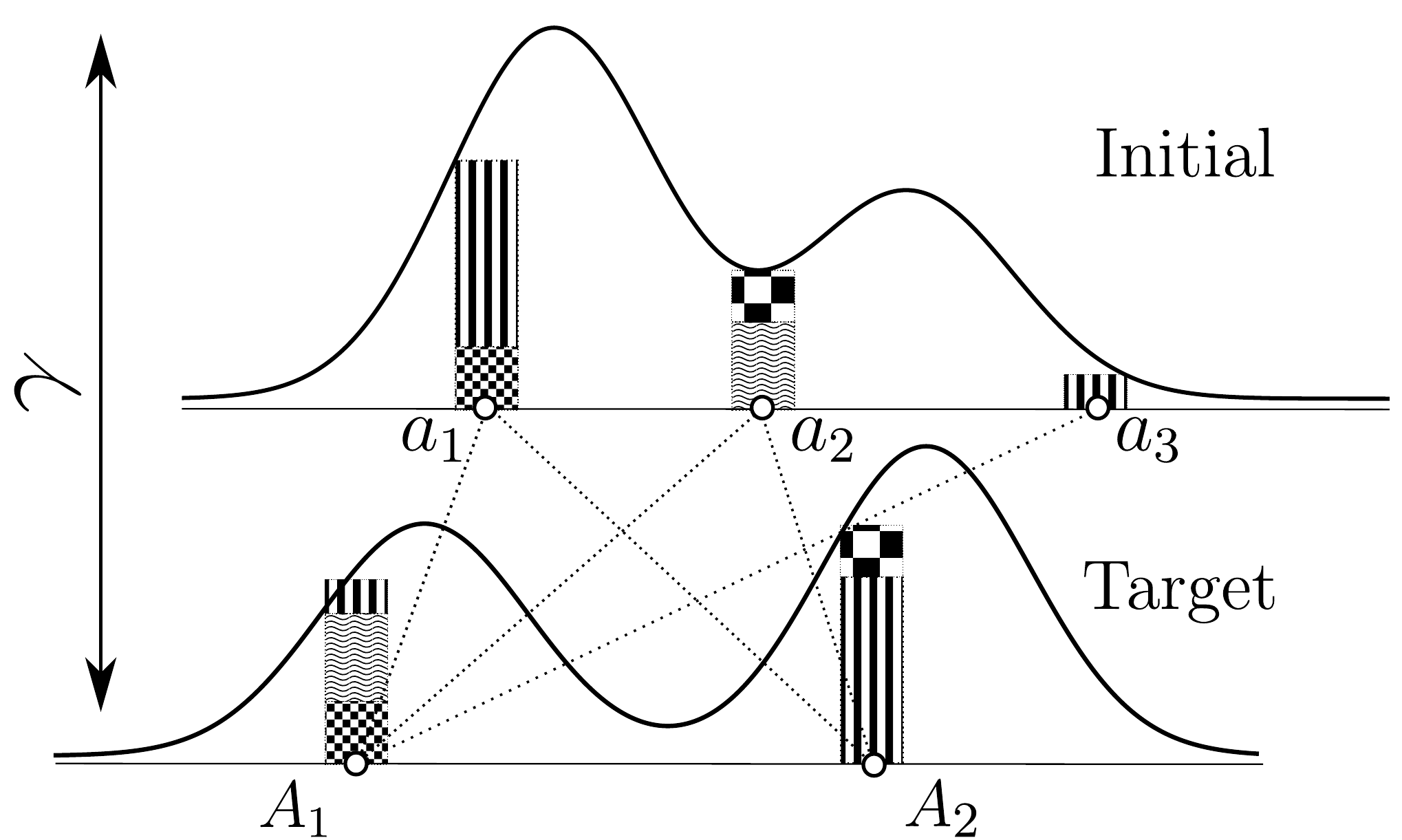}
\caption{Illustrating some transport plan (not necessarily the optimal) 
between some initial and target distributions. 
The patterned boxes indicate how an amount of distribution is redistributed from an initial set of points $a_1,a_2,$ and $a_3$ to points $A_1$ and $A_2$.}
\label{transport_plan}
\end{figure}

As illustrated in Fig.~\ref{transport_plan}, the Wasserstein metric is the minimal amount of work needed to transform one distribution into another; work being defined as the amount of distribution times the distance it has to be transported. There are many different ways of transporting an amount of distribution from a region $x$ of $p$ into a region $y$ of $q$. The set of all possible transport plans to move $p$ into $q$ is denoted by $\Gamma$. Hence, computing the distance between two distributions can be formulated as an optimization problem where the aim is to find that transport plan $\gamma\in\Gamma$ such that the total amount of work is minimal. The Wasserstein~\cite{Wasserstein1969} metric is expressed as
\begin{equation}
    W_1(p, q)=\inf_{\gamma \in \Gamma} \int_{\mathbb{R}^d\times \mathbb{R}^d} \mathrm{d}\gamma(x,y)\; \left|x-y\right| \,= \,\int_{\mathbb{R}}\mathrm{d}t\; \left|P(t)-Q(t)\right|,
    \label{eqn:wasserstein_metric}
\end{equation}
where the right-hand side equation was shown to hold in Ref.~\cite{Ramdas2017OnTests} with $P$ and $Q$ being the cumulative distribution functions of $p$ and $q$. The vectorized two-dimensional Coulomb matrix representation of a molecule can be used in Eq.~(\ref{eqn:wasserstein_metric}). From now on, we set the $L_1$ and $L_2$ based kernel functions $k$ to be the well known Laplacian and Gaussian kernel, respectively. Additionally, we define the Wasserstein-based kernel as
\begin{equation}
    k(\mathbf{x}, \mathbf{y})=e^{-\alpha W_1(\mathbf{x}, \mathbf{y})}.
    \label{eqn:wasserstein_kernel}
\end{equation}
Here, we note that other kernel functions or representations could be used in combination with the Wasserstein metric just as well. 

All QML models of atomization energies of QM9 molecules~\cite{QM9} were trained using KRR with 5-fold cross-validation for hyperparameter optimization, and tested on ~2000 out-of-sample molecules. 

\section{Results and discussion}

\begin{figure}[tb]
\centering
\includegraphics[width=\linewidth]{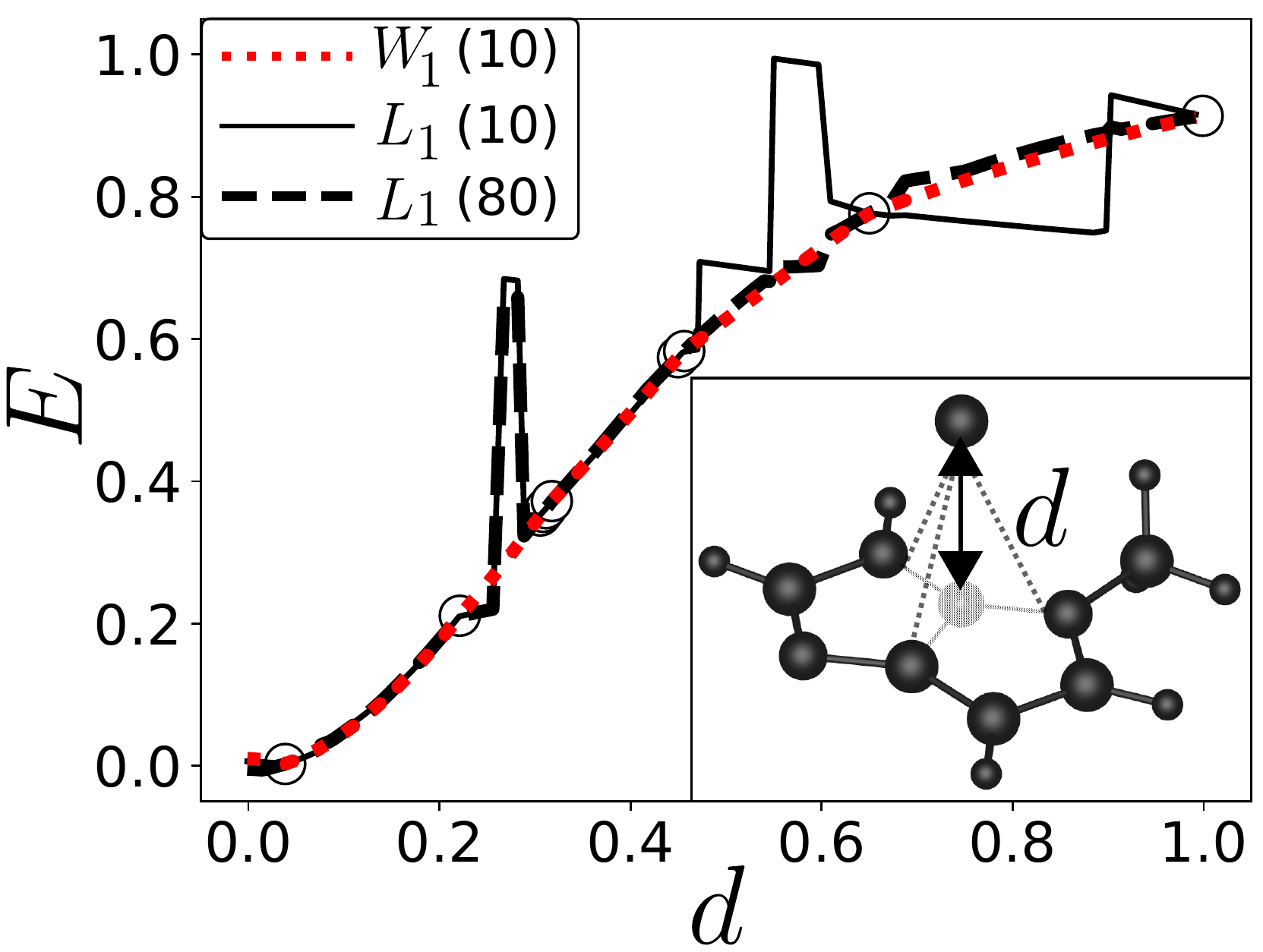}
\caption{Normalized model Lennard-Jones potential energy $E$ as a function of normalized displacement $d$ (in steps of 0.01 {\AA}) of an atom from its equilibrium position in a molecule as shown in the inset.  Randomly selected training data points are indicated as open circles. CM based QML predictions with the Wasserstein kernel $W_1$ trained on ten data points (dotted line) are smooth, while discontinuities are visible for the $L_1$ norm for small (solid) and large (dashed) training sets.}
\label{fig:modelmolecule}
\end{figure}

First, we illustrate the issue of smoothness by subjecting an innocent organic molecule to drastic distortions. More specifically, and as shown in Fig.~\ref{fig:modelmolecule}, consider the energy $E$ as a function of continuous displacement $d$ (in steps of 0.01 {\AA}) of some central carbon atom along an axis orthogonal to the molecular plane. Corresponding Lennard-Jones potential energies are smooth, while CM based QML model predictions (after training on 10 instances drawn at random) using $L_1$ are discontinuous. Even after increasing the training set for the $L_1$ model to 80 instances, the discontinuity at $d\approx 0.3$ is retained, indicating that lack of learning. As pointed out above, the main benefit of Eq.~(\ref{eqn:wasserstein_kernel})  is that $W_1$ is invariant under permutations of row- and column-indices of the adjacency matrix. When applied to the model system from Fig.~\ref{fig:modelmolecule}, which showed the typical indexing problem with the Laplacian kernel and $L_1$-norm, the $W_1$ metric yields smooth energies as a function of displacement. Two aspects are noteworthy: First, it is apparent that the predictions are smooth (differentiable) across the full range of displacements and do not exhibit any discontinuities that affected the standard kernels in combination with the sorted Coulomb matrix. Second, a training data set including only 10 reference points is sufficient to give accurate results. These numerical results clearly indicate that the use of the Wasserstein metric in the kernel cures the indexing problem and alleviates the associated low prediction quality and data efficiency.   

\begin{figure}[tb]
\centering
\includegraphics[width=\linewidth]{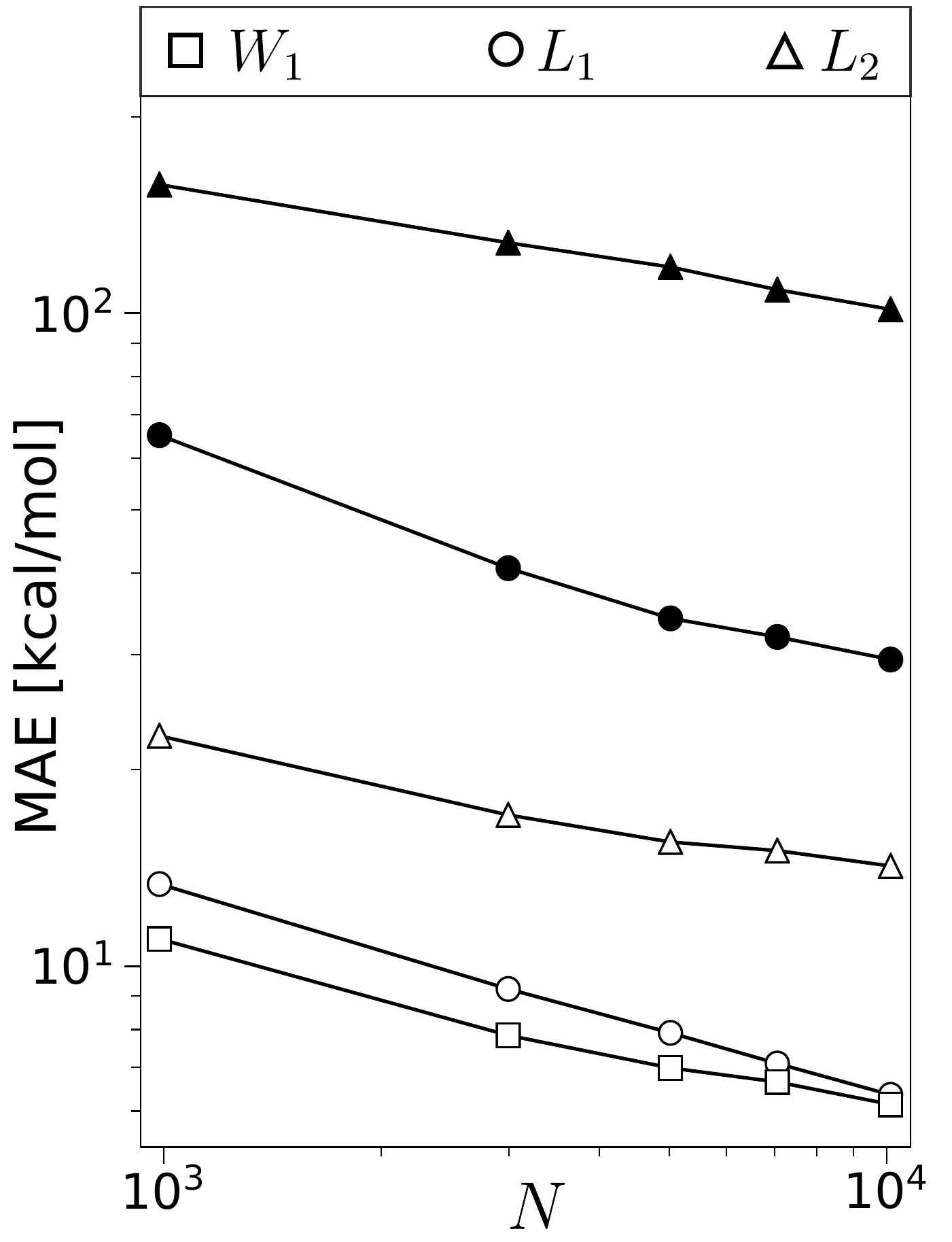}
\caption{Learning curves for DFT/B3LYP/6-31G(2df,p) atomization energies of QM9 molecules for various kernels and representations. Circles and triangles indicate the MAE obtained with $L_1$ and $L_2$ based kernels, respectively, using randomized (filled symbols) and sorted (open symbols) Coulomb matrix representation. The Wasserstein metric is permutation invariant and the MAE obtained with it is given by squares. }
\label{fig:QM9set}
\end{figure}

To further investigate the performance of Wasserstein based kernels (Eq.~(\ref{eqn:wasserstein_kernel}) in QML models, we have turned to the classic benchmark of atomization energies of organic molecules in the QM9 data set~\cite{QM9}. After training on up to 10k molecules, randomly sampled from the entire QM9 data set, learning curves are presented in Fig.~\ref{fig:QM9set}. Results for $L_2$ and $L_1$ based kernels are in line with observations made in Ref.~\cite{Hansen2013AssessmentEnergies}: For sorted as well as for randomly indexed CMs the $L_1$ based QML model exhibits lower off-sets than the $L_2$ based model. Not surprisingly, use of sorted CMs also results in smaller off-sets than for random CMs. More interestingly, however, the $W_1$ based metric results in the same learning curve after being applied to random as well as to sorted CMs.  Its overall learning curve off-set and slope indicates same (even slightly better) 
performance as the $L_1$ norm applied to the sorted CM, reaching
$\sim$6 kcal/mol after training on 10k instances. 

\section{Conclusions}
Considering the findings for the continuous atomic displacement as well as the QM9 molecules, the Wasserstein metric enables the generation of QML models which achieve (a) data-efficient learning and (b) smooth target function estimates. While all our numerical evidence has relied on the CM, we emphasize that the observed solution of the indexing problem and the simultaneously improvement of the predictions by using the Wasserstein kernel is not inherently specific to the CM representation. In fact, the Wasserstein metric can readily be applied to QML models relying on any graph-based representation. This could be  particularly relevant in the context of recent work on learning force-fields or electronic properties, relying on inverse distances rather than the CM representation~\cite{chmiela2018towards,chmiela2019sgdml,westermayr2019neural,westermayr2019machine}

To summarize, we have presented  the Wasserstein metric as an index-invariant way to measure distances between molecular graph-based representations. Our numerical findings indicate that the resulting QML models combine smoothness with data efficiency in learning. Future work will explore the various possible combinations of kernel functions, Wasserstein metric, and representations other than the CM. 

\section*{Acknowledgement}
O.A.v.L. acknowledges funding from the Swiss National Science foundation (No.~407540\_167186 NFP 75 Big Data) and from the European Research Council (ERC-CoG grant QML). This work was partly supported by the NCCR MARVEL, funded by the Swiss National Science Foundation. B. B. acknowledges support by the Innovational Research Incentives Scheme Vidi of the Netherlands Organisation for Scientific Research (NWO) with project number 723.016.002. Partial funding is also provided by NWO and the Netherlands eScience Center for funding through project number 027.017.G15, within the Joint CSER and eScience program for Energy Research (JCER 2017).


\end{document}